\documentclass[11pt]{article}
\usepackage[letterpaper,margin=1in]{geometry}
\usepackage[T1]{fontenc}
\usepackage{lmodern}
\usepackage{amsmath,amssymb,amsthm,bm,booktabs,graphicx,threeparttable,mathtools}
\usepackage[font=small,labelfont=bf,labelsep=period,skip=6pt]{caption}
\usepackage[round,authoryear]{natbib}
\usepackage{microtype}
\usepackage{setspace}
\usepackage{placeins}
\usepackage{enumitem}
\usepackage[hidelinks]{hyperref}
\hypersetup{
  pdftitle={Identifying the Discount Factor in Dynamic Discrete Choice with Backward Autoregressions},
  pdfauthor={Haojie Liu, Zihan Lin}
}
\newtheorem{theorem}{Theorem}
\newtheorem{proposition}{Proposition}
\theoremstyle{definition}
\newtheorem{assumption}{Assumption}
\newcommand{\R}{\mathbb R}
\newcommand{\E}{\mathbb E}

\newcommand{\col}{\operatorname{col}}
\newcommand{\row}{\operatorname{row}}
\newcommand{\B}{\mathcal B}
\newcommand{\Omat}{\mathcal O}
\newcommand{\exhibitnote}[1]{%
  \par\smallskip
  \begin{minipage}{.95\linewidth}\raggedright\footnotesize #1\end{minipage}%
}
\title{Identifying the Discount Factor in Dynamic Discrete Choice\\
with Backward Autoregressions}
\author{%
  \begin{tabular}{@{}c@{\hspace{2.6em}}c@{}}
  Haojie Liu & Zihan Lin\\[2pt]
  \small\href{mailto:hliu332@ucr.edu}{\nolinkurl{hliu332@ucr.edu}} &
  \small\href{mailto:zlin169@ucr.edu}{\nolinkurl{zlin169@ucr.edu}}\\[5pt]
  \multicolumn{2}{c}{\small Department of Economics, University of California, Riverside}
  \end{tabular}%
}
\date{}

\begin{document}
\maketitle
\begin{abstract}
We study how to identify the discount factor in dynamic discrete choice models when the data cover only a finite window. The continuation value beyond the window is unrestricted. If payoffs and transitions are constant over time, differencing the log odds of choice removes flow payoffs. In a model with two states, three dates can then identify the discount factor in closed form. With more states, the transition matrices determine how many later dates are needed to eliminate the continuation values. The resulting backward autoregression and a check on the last dates characterize the identified set. With known transitions and regular identification, minimum distance using all of these conditions and the inverse of their covariance matrix has the same asymptotic variance as maximum likelihood based on solving the model. In simulations, dropping equations near the end of the window raises error. A likelihood search with one autoregression start and one fixed start saves a median of 63\% of computation time relative to four fixed starts and attains the same likelihood within numerical tolerance.
\end{abstract}

\section{Introduction}

A firm pays to repair a machine today in exchange for a lower chance of operating it in poor condition tomorrow. A firm with several years left on its contract can benefit from a repair for longer than a firm whose contract is about to end. The firms can therefore make different choices at the same machine condition and repair cost. We ask when such differences identify the discount factor if the value of the machine after the last observed date is unknown.

At one date, the repair probability reflects both its immediate payoff and its effect on future payoffs. Comparing two dates at the same machine condition removes the immediate payoff. That comparison still gives only the product of the discount factor and a change in continuation values. In an example with two states, a third date gives a second comparison that eliminates the unknown value change and leaves an explicit formula for the discount factor.

\citet[][Section 3]{an2021} use the same two equations to study a different parameter. They fix the discount factor, difference the log choice odds, and substitute the value recursion to identify subjective transition beliefs. We supply the transition matrix for each action and solve for the discount factor. When the difference between the two transition matrices has full column rank after we express values relative to a reference state, their substitution gives our equation of order one. With deficient rank, the choice probabilities may reveal only some combinations of value changes. We use the transition under the reference action to determine which further combinations can be recovered from later dates.

This calculation gives a backward autoregression in observed choice probabilities for each admissible order. The shortest order depends on the transitions. An equation of order two links the change in log odds today to quantities at two later dates, with coefficients involving the first and second powers of the discount factor. The last few dates require a separate check because the later observations needed for those equations lie outside the window. We show that the recurrence and this check give the sharp identified set. For each discount factor in that set, we construct flow payoffs and a continuation value that reproduce the choices.

The argument starts from logit inversion \citep{hotz1993}. Time variation is also used by \citet{tsubota2026} to study hyperbolic discounting. Other identification arguments restrict the payoffs or how they enter the model \citep{magnac2002,abbring2020,komarova2018}. \citet{am2020} study identification of payoffs and counterfactual choices in short panels. \citet{chou2025} use conditional means in nonstationary models and can avoid estimating transition probabilities. In our setting, the transition matrices determine how many dates are needed for the recurrence.

With known transitions, the same conditions also give an estimator based on minimum distance. When the discount factor is regularly identified, using all the conditions and weighting them by the inverse of their covariance matrix gives the same asymptotic variance as maximum likelihood based on solving the model.

In the simulations, the population choice probabilities first identify the discount factor with three, four, and five dates in the designs of order one, two, and three, respectively. In the design of order one with more information about discounting, the root mean squared error rises from 0.055 to 0.305 when we fit a longer recurrence that drops equations near the end of the window. Adding those equations recovers the shorter fit. We then compare minimum distance with likelihood under known and estimated transitions, and examine how much the confidence sets rule out under weak identification. A comparison with the likelihood of \citet{tsubota2026} examines what changes when we estimate both the factor for present bias and the discount factor for the long run.

\section{Dynamic choice and time variation}
\label{sec:model}

\subsection{Decision problem}

The agent observes one of $K\ge2$ states and chooses $a=0$ or $a=1$. In the repair example, action 0 keeps the machine in service and action 1 repairs it. Action $a$ yields $u_a(s)+\epsilon_t(a)$ in state $s$ and determines the distribution of next period's state. The agent discounts future payoffs geometrically with factor $\beta_0\in(0,1)$.

We observe dates $1,\ldots,H$, which may cover only part of the decision problem. The continuation values $V_{H+1}$ are unrestricted.

\begin{assumption}[Time invariance]
For each action $a$, the flow payoff $u_a(s)$ and transition probabilities $P_a(s,s')$ are constant over the observed window. The agent uses those transition probabilities when evaluating future payoffs.
\end{assumption}

\begin{assumption}[Private payoff shocks]
The shocks $\epsilon_t(a)$ follow a type I extreme value distribution with unit scale and are independent across actions and dates. The next state is independent of the shocks conditional on the current state and action.
\end{assumption}

\subsection{Choice probabilities and the Bellman equation}

Let $V_t(s)$ be the value at the start of date $t$, before the private shocks are drawn. The value of choosing action $a$ before its shock is
\[
Q_{a,t}(s)=u_a(s)+\beta_0\sum_{s'}P_a(s,s')V_{t+1}(s').
\]
The agent chooses the action with the larger $Q_{a,t}(s)+\epsilon_t(a)$. Let $p_t(s)$ be the probability of action 1 given state $s$. Under the usual logit location normalization, the log odds equal the difference between the values of the two actions.
\begin{equation}
L_t=\log\frac{p_t}{1-p_t}
   = Q_{1,t}-Q_{0,t}
   = u_1-u_0+\beta_0(P_1-P_0)V_{t+1}.
\label{eq:logit}
\end{equation}
The vectors $p_t$, $L_t$ and $Q_{a,t}$ have one entry per state. We take logarithms and ratios componentwise and assume $0<p_t(s)<1$ at every observed state and date.

Averaging the value of the optimal choice over the shocks gives $V_t=Q_{0,t}+\log(1+\exp L_t)$. Since $\log(1+\exp L_t)=-\log(1-p_t)$, the Bellman equation for $V_t$ can be written in terms of the observed choice probabilities.
\begin{equation}
V_t=u_0-\log(1-p_t)+\beta_0P_0V_{t+1}.
\label{eq:bell}
\end{equation}
When the agent almost always chooses action 0, the adjustment $-\log(1-p_t)$ is small. Equation~\eqref{eq:logit} tells us which combination of continuation values enters the log odds. Equation~\eqref{eq:bell} tells us how those values are linked across dates.

\subsection{What differencing removes}

Subtract the log odds at date $t+1$ from those at date $t$ for the same state. The flow payoff difference cancels.
\begin{equation}
z_t:=L_t-L_{t+1}
=\beta_0(P_1-P_0)(V_{t+1}-V_{t+2}).
\label{eq:logitdiff}
\end{equation}
\citet[][equations (5)--(6)]{an2021} use the same differencing step. Equation~\eqref{eq:logitdiff} still contains two unknowns, $\beta_0$ and the change in continuation values. Differencing the Bellman equation removes $u_0$ and links that value change to the next one.

\section{Identification from consecutive dates}
\label{sec:identification}

\subsection{Two states and three observed dates}
\label{sec:two-state}

Suppose the machine can be in good condition or worn. Let $v_t$ be the value of a worn machine relative to one in good condition, and let $d(s)=P_1(s,\mathrm{worn})-P_0(s,\mathrm{worn})$ be the effect of repair on the probability of wear next period. The two state probabilities sum to one, so repair changes the expected continuation value by $d(s)v_{t+1}$. Equation~\eqref{eq:logitdiff} reduces to
\begin{equation}
z_t(s)=\beta_0\,d(s)\,(v_{t+1}-v_{t+2}).
\label{eq:two-state-logit}
\end{equation}
If $d(s)\ne0$, the log odds tell us the product of the discount factor and the value change. With two dates, we can match that product at any discount factor by adjusting the value change. Appendix~\ref{app:boundary} gives the general compatibility check.

A third date lets us difference the Bellman equation between dates 2 and 3. Let $h_t$ be the difference between the worn and good states in $-\log(1-p_t)$. Let $\rho=P_0(\mathrm{worn},\mathrm{worn})-P_0(\mathrm{good},\mathrm{worn})$ measure how much current wear raises the probability of future wear when the firm keeps the machine. Then
\[
v_2-v_3=(h_2-h_3)+\beta_0\rho(v_3-v_4).
\]
The flow payoff under action 0 has canceled. The log odds at dates 2 and 3 give $z_2(s)=\beta_0d(s)(v_3-v_4)$. We can now substitute for $v_2-v_3$ and then for the term involving $v_3-v_4$ in the equation for $z_1(s)$.
\begin{equation}
z_1(s)=\beta_0\big[d(s)(h_2-h_3)+\rho z_2(s)\big].
\label{eq:two-state-beta}
\end{equation}
The bracket contains only choice probabilities and transitions. If it is nonzero in any state, divide $z_1(s)$ by the bracket to recover $\beta_0$. The calculation leaves $v_4$ unrestricted.

\begin{table}[htbp]
\centering
\caption{Repair probabilities in the example with two states}
\label{tab:two-state-choices}
\begin{tabular}{crr}
\toprule
Observed date & Good condition & Worn condition \\
\midrule
1 & 0.2515 & 0.5767 \\
2 & 0.2719 & 0.6023 \\
3 & 0.3363 & 0.6726 \\
\bottomrule
\end{tabular}

\end{table}

In Table~\ref{tab:two-state-choices}, a machine left unrepaired is worn next period with probability 0.25 from good condition and 0.75 from worn condition. Repair lowers both probabilities by 0.20, so $\rho=0.50$ and $d(s)=-0.20$. The first two changes in log odds in the good state are $-0.10570$ and $-0.30508$ to five decimal places. With $h_2-h_3=-0.10205$, the bracket in \eqref{eq:two-state-beta} is $-0.13213$. The ratio gives $\beta_0=0.80$ in both states. Appendix~\ref{app:simulations} lists the inputs used to solve the model by backward induction and generate these population probabilities.

\subsection{Several states and the direct inversion benchmark}
\label{sec:inversion}

With more states, $v_t$ is the vector of values relative to the first state. Form $D$ by deleting the first column of $P_1-P_0$. For $P$, subtract the first row of $P_0$ from each of its other rows and then drop the first row and column. The vector $h_t$ is $-\log(1-p_t)$ measured relative to the first state. The two equations in levels are
\begin{equation}
L_t=b+\beta_0 Dv_{t+1},\qquad v_t=c+h_t+\beta_0Pv_{t+1},
\label{eq:reduced}
\end{equation}
where $b$ is the flow payoff difference between actions and $c$ is the vector of flow payoffs for action 0 relative to the first state. In addition to $z_t=L_t-L_{t+1}$, let $g_t=h_t-h_{t+1}$ and $w_t=v_t-v_{t+1}$. Taking time differences removes $b$ and $c$.
\begin{equation}
z_t=\beta_0Dw_{t+1},\qquad w_t=g_t+\beta_0Pw_{t+1}.
\label{eq:ss}
\end{equation}
In the derivation with two states, $w_t$, $g_t$, and the relative transition $P=\rho$ are scalars.

Changing the reference state and reversing the time difference maps equations (6) and (8) of \citet{an2021} to \eqref{eq:ss}. They use an invertible square block of the transition contrast to eliminate continuation values, treating discounting as given. Here the transitions are given. If $D$ has full column rank, its Moore--Penrose inverse $D^+$ is a left inverse, and the first equation gives $\beta_0 w_{t+2}=D^+z_{t+1}$. Substituting into the second equation gives
\begin{equation}
z_t=\beta_0\big(Dg_{t+1}+DPD^+z_{t+1}\big).
\label{eq:direct-inversion}
\end{equation}
If we keep the same invertible rows as \citet{an2021}, this is their equation (10) premultiplied by the transition contrast and reindexed. With that contrast given, it is an equation of order one in the discount factor.

\subsection{Elimination when direct inversion is unavailable}
\label{sec:two-step}

Observing $Dw$ determines all of $w$ if $D$ has full column rank. To form the recurrence, it suffices to know $DPw$. If $DP=C_0D$ for a matrix $C_0$, we can compute $DPw=C_0Dw$ even when several $w$ give the same $Dw$. Substitution into \eqref{eq:ss} gives \eqref{eq:direct-inversion} with $C_0$ in place of $DPD^+$.

With three states, suppose the choice reveals the sum of the two relative values. If the reference action carries the two values forward at different rates, their sum today cannot tell us their weighted sum tomorrow. The extra weighted sum supplies a second combination. With two relative values, the characteristic equation of $P$ makes the third transition block a combination of the first two.

If $DPw$ cannot be computed from $Dw$, a further date can help. Write $y_t=\beta_0 w_{t+1}$ so that
\begin{equation}
z_t=Dy_t,\qquad y_t=\beta_0 g_{t+1}+\beta_0Py_{t+1}.
\label{eq:y}
\end{equation}
The second equation applies whenever both $y_t$ and $y_{t+1}$ are in the window. Iterating it twice gives
\[
z_t=\beta_0Dg_{t+1}+\beta_0^2DPg_{t+2}+\beta_0^2DP^2y_{t+2}.
\]
The later differences already reveal $Dy_{t+2}=z_{t+2}$ and $\beta_0DPy_{t+2}=z_{t+1}-\beta_0Dg_{t+2}$. If $DP^2=C_0D+C_1DP$, these two combinations suffice to substitute for the last term. The result is
\begin{equation}
z_t=\beta_0(Dg_{t+1}+C_1z_{t+1})
 +\beta_0^2\big((DP-C_1D)g_{t+2}+C_0z_{t+2}\big).
\label{eq:order-two}
\end{equation}
We use four consecutive dates of choice probabilities. The two powers in \eqref{eq:order-two} belong to the same discount factor. Expressing the third transition block in terms of the first two removes the remaining unknown value vector.

\subsection{General order and the identified set}

Let $n=H-1$ be the number of date differences. An order $1\le q<n$ is admissible if the rows of $DP^q$ are linear combinations of the rows already present in $D,DP,\ldots,DP^{q-1}$. Equivalently, there are matrices $C_j$ such that
\begin{equation}
DP^q=\sum_{j=0}^{q-1}C_jDP^j.
\label{eq:C}
\end{equation}
We can find the first such order by checking when the rank of the stacked blocks stops increasing. Multiplying by $P$ then keeps every later block in the same row space. With $K$ states and enough observed dates, the characteristic equation of $P$ ensures closure by order $K-1$. At an admissible order, the same substitution yields
\begin{equation}
z_t=\sum_{j=1}^{q}\beta_0^j a_{j,t},\qquad t=1,\ldots,n-q,
\label{eq:ar}
\end{equation}
where $a_{j,t}$ is a known linear combination of $z_{t+j}$ and $g_{t+j}$. Appendix~\ref{app:model} gives the coefficients.

The local equations leave the last $q$ differences to check. For example, at order two a candidate $\beta$ must admit a single boundary vector $y_n$ satisfying
\[
z_n=Dy_n,\qquad z_{n-1}-\beta Dg_n=\beta DP y_n.
\]
For general $q$, one vector must satisfy all $q$ boundary equations. For a given $\beta$, they form the linear system in Appendix~\ref{app:boundary}.

\begin{theorem}[Representation by a finite recurrence and identification]
\label{thm:ident}
Under the model in Section~\ref{sec:model}, let $q$ be admissible.
\begin{enumerate}[label=(\roman*),nosep]
\item Any discount factor that rationalizes the observed choice probabilities satisfies \eqref{eq:ar}.
\item A candidate discount factor rationalizes those probabilities if it satisfies \eqref{eq:ar} and the conditions for the final dates. Together these conditions characterize the sharp identified set.
\item A unique common root of \eqref{eq:ar} in the parameter interval suffices for point identification. One sufficient condition is that the $q$ coefficient columns formed by stacking $a_{j,t}$ over states and dates are linearly independent.
\end{enumerate}
\end{theorem}

To prove part (ii), we use the boundary vector to recover the earlier differences and construct payoffs that are constant over time and a continuation value that reproduce the observed probabilities. The boundary equations can rule out some roots of the recurrence. At the other extreme, if the choice probabilities are constant across dates in every state, $z_t$ and $g_t$ are zero and every discount factor is compatible. Thus an admissible order determines when we can write an equation. Whether it identifies discounting depends on how choices change across dates.

\section{Estimation and inference}
\label{sec:estimation}

\subsection{Estimating the discount factor by minimum distance}

In the example with two states in Section~\ref{sec:two-state}, a single ratio recovers the discount factor. Once we estimate the choice probabilities, ratios from different states and dates will generally differ. For $q>1$, the restrictions also involve higher powers of the same parameter. We fit all the restrictions by minimum distance.

Fix an admissible order $q$ and a parameter interval $\mathcal B$ within $(0,1)$. Estimate the choice probabilities from frequencies by state and date. Use supplied transition matrices or estimate them in the first stage. At each candidate $\beta$, let $\hat m(\beta)$ collect the sample residuals from \eqref{eq:ar} and any boundary or output restrictions included in the fit.
\begin{equation}
\hat\beta\in\arg\min_{\beta\in\mathcal B}
        \hat m(\beta)^\top \hat W(\beta)\hat m(\beta).
\label{eq:est}
\end{equation}
Neighboring equations use some of the same estimated choice probabilities, so their residuals are correlated. In the simulations, $\hat W(\beta)$ is the inverse of the estimated moment covariance on its nonredundant space. We recompute this weight at each candidate, giving the continuously updated estimator. Once the choice probabilities and transitions are available, the search is over $\beta$ alone.

\subsection{Likelihood with and without present bias}
\label{sec:likelihood-benchmarks}

Minimum distance fits the restrictions on changes in log odds. Maximum likelihood fits choices in levels by solving the model \citep{rust1987}. With the transition matrices given, its objective is the conditional choice log likelihood
\begin{equation}
\ell(\theta)=\sum_{t,s}\big[A_{ts}\log p_t(s;\theta)
 +(N_{ts}-A_{ts})\log(1-p_t(s;\theta))\big],
\label{eq:choice-likelihood}
\end{equation}
where $N_{ts}$ is the number of visits to a state and date and $A_{ts}$ is the number of times action 1 is chosen. At each trial vector $\theta$, the geometric MLE computes the choice probabilities by backward induction and searches jointly over the discount factor, the flow payoffs and the terminal values. Both the AR estimator and the geometric MLE allow the value beyond the observed window to be unknown.

\citet{tsubota2026} also uses variation in choices over time. His identification argument imposes payoff equalities across selected combinations of states and actions and rank conditions on the resulting equations. His model places weights $\alpha\delta,\alpha\delta^2,\ldots$ on future payoffs. The factor $\alpha$ captures present bias, and $\delta$ governs the rate of discounting in the long run. He estimates both by maximum likelihood, computing choice probabilities by backward induction with a value of zero after the terminal date.

For a comparison on the same observed window, we adapt Tsubota's likelihood to use the same known transitions and free terminal values as our geometric MLE. With $\alpha=1$, its recursion and objective reduce to the geometric MLE at $\delta=\beta$. Estimating $\alpha$ gives a distinct third fit while keeping the likelihood in \eqref{eq:choice-likelihood} and the observed sample the same. Appendix~\ref{app:present-bias} gives the recursion and the simulation protocol. Appendix~\ref{app:likelihood-starts} separately examines whether minimum distance can help initialize the geometric likelihood search.

\subsection{Asymptotic properties}

Let $M$ be the number of independent paths or, under recurrent sampling, the number of repeated observation cycles. The horizon and order stay fixed as $M$ grows. Appendix~\ref{app:statistics} states the conditions for the first stage, identification and weighting.

\begin{theorem}[Consistency]\label{thm:cons}
Under conditions (i)--(iv) in Appendix~\ref{app:statistics}, any measurable global minimizer of \eqref{eq:est} converges in probability to $\beta_0$.
\end{theorem}

\begin{theorem}[Asymptotic normality]\label{thm:normal}
Under conditions (i)--(vii), $\sqrt M(\hat\beta-\beta_0)$ is asymptotically normal with mean zero and variance given in \eqref{eq:variance}.
\end{theorem}

\begin{proposition}[Same asymptotic variance as geometric MLE]\label{prop:efficiency}
Suppose the transition matrices are known, the paths are independent, and $\beta_0$ is interior and regularly identified. The consistent estimators based on minimum distance and geometric maximum likelihood have the same asymptotic variance for $\beta$ when minimum distance uses all restrictions across the window, with redundant equations removed and the inverse of their covariance matrix as the weight.
\end{proposition}

At a given discount factor, the restrictions across the window describe exactly the choice probabilities that the likelihood can fit by varying the payoffs and terminal values. This is why the two procedures retain the same local information about discounting. When the local recurrence omits some of those restrictions, the result applies to the fit that also uses the missing conditions. Appendix~\ref{app:efficiency} gives the proof.

Choices within an independent path can still be dependent. For recurrent data, dependence across cycles enters the covariance of the estimates from the first stage. When the data give little information about $\beta$, we form a confidence set by testing each candidate value \citep{stock2000}. Appendix~\ref{app:model} gives the test statistic.

\section{Monte Carlo experiments}
\label{sec:experiments}

\subsection{Population identification and simulation design}

In the example with two states, three observed dates provide the two changes in log odds needed to recover the discount factor. We first ask how many dates each of the simulated models needs in the population.

We simulate seven dates of binary choices with four states and $\beta_0=0.8$. The three transition systems have minimum orders one, two and three. In each system we scale the terminal payoff to match two levels of local likelihood information about the discount factor after accounting for unknown payoffs. The higher level has ten times the information per path of the lower level. The likelihood can have a different shape away from the true value. We treat the terminal payoffs as unknown and supply the transition matrices except in Section~\ref{sec:feasible-experiment}.

We add one date at a time to the same population choice path. The designs of order one, two and three first identify $\beta=0.80$ with three, four and five dates, respectively. With one fewer date, any discount factor in $(0,1)$ can reproduce the choices. This pattern holds at both information levels. The minimum order determines when the first local equation becomes available. The population choice probabilities determine its roots.

We start with 10,000 independent paths, then use multiples of 50,000 from 50,000 to one million. We run 200 replications at each of the 21 sizes and fit the estimators to the same draws. Appendix~\ref{app:simulations} gives the calibration and numerical details.

\subsection{What the final dates contribute}

When the shortest recurrence has order one or two, a recurrence of order three leaves more dates at the end of the window without their own equation. We fit the longer recurrence on the same paths with and without those final equations. At the higher information level and one million paths, adding the equations cuts root mean squared error (RMSE) from 0.305 to 0.055 when the shortest order is one, and from 0.077 to 0.057 when it is two.

Once we put the final equations back, the longer fit agrees with the shortest fit to numerical precision at every sample size. In these designs the two fits use the same conditions written in different ways. Weighting by their covariance gives the same objective.

\subsection{AR and geometric MLE}

\begin{figure}[!htbp]
\centering
\includegraphics[width=.97\linewidth]{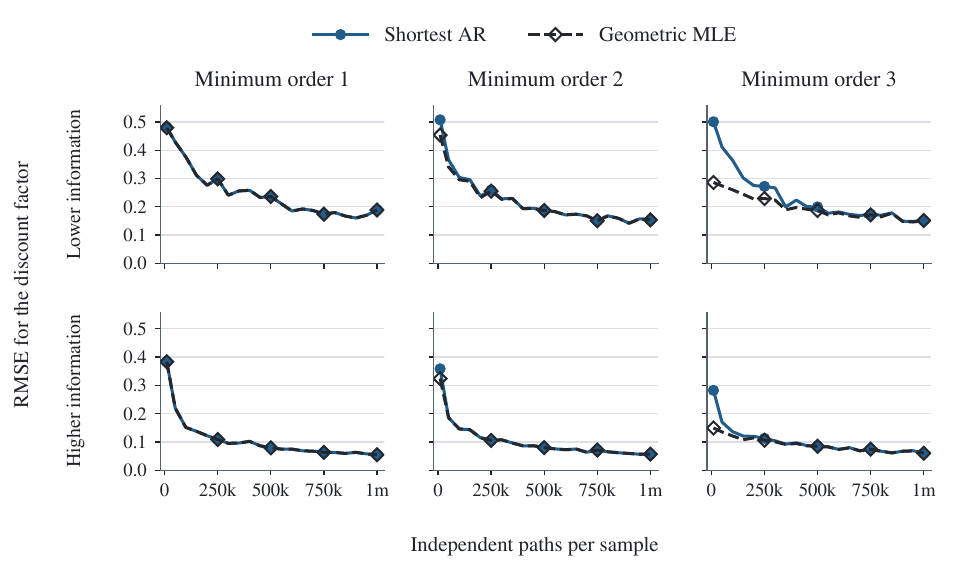}
\caption{RMSE for the discount factor at matched information levels}
\label{fig:matched}
\exhibitnote{
Each point uses 200 matched samples. Shortest AR is the shortest admissible backward autoregression. The rows differ in local likelihood information about the discount factor. Lines join the simulated sample sizes.
}
\end{figure}

Figure~\ref{fig:matched} shows that the shortest recurrence closely tracks geometric MLE in the design of order one. The largest gaps occur at order three. At the lower information level and 100,000 paths, RMSE is 0.364 for AR and 0.260 for MLE. At one million paths, both round to 0.152. At the higher information level and one million paths, the RMSEs of both methods range from 0.055 to 0.060 across all three orders. The shortest recurrence uses the full set of restrictions in these designs, as required by Proposition~\ref{prop:efficiency}. The proposition concerns the limit as the sample grows. The gaps at smaller sample sizes remain large in some designs.

The AR estimate can also serve as a starting value for maximum likelihood. On 1,800 matched draws, we compare a search with one AR start and one fixed start against a benchmark with four fixed starts. The median ratio of the time for the two starts to the benchmark time is 0.37, including the time to compute the AR estimate. The fitted log likelihoods agree within 0.01 on every draw. Appendix~\ref{app:likelihood-starts} reports the timing protocol and the results from either start alone.

\subsection{Estimating the transition probabilities}
\label{sec:feasible-experiment}

We return to the design of order three and estimate the transition probabilities from the same paths used to estimate choice probabilities. We pool transitions across dates within each state and action. AR uses the restrictions across the entire window and accounts for transition estimation in its weighting matrix. Geometric MLE plugs the same transition estimates into the choice likelihood. We compare both fits with the fits that use known transitions on the same draws.

The effect of estimating transitions varies across the two information levels (Table~\ref{tab:estimated-transitions}). At the higher information level and one million paths, AR RMSE changes from 0.060 to 0.062, and geometric MLE RMSE rises from 0.060 to 0.097. Much of the latter increase comes from upward bias, which rises from 0.013 to 0.067. At the lower information level and 100,000 paths, geometric MLE with estimated transitions has lower RMSE than AR, 0.284 versus 0.392.

\begin{table}[!htbp]
\centering
\begin{threeparttable}
\caption{RMSE with known and estimated transitions}
\label{tab:estimated-transitions}
\begin{tabular}{r rr rr}
\toprule
 & \multicolumn{2}{c}{Known transitions} & \multicolumn{2}{c}{Estimated transitions} \\
\cmidrule(lr){2-3}\cmidrule(lr){4-5}
Paths & AR & MLE & AR & MLE \\
\midrule
\multicolumn{5}{l}{\textit{Lower information}} \\
10,000 & 0.501 & 0.286 & 0.534 & 0.401 \\
100,000 & 0.364 & 0.260 & 0.392 & 0.284 \\
1,000,000 & 0.152 & 0.152 & 0.151 & 0.137 \\
\addlinespace
\multicolumn{5}{l}{\textit{Higher information}} \\
10,000 & 0.282 & 0.149 & 0.178 & 0.171 \\
100,000 & 0.136 & 0.121 & 0.139 & 0.156 \\
1,000,000 & 0.060 & 0.060 & 0.062 & 0.097 \\
\bottomrule
\end{tabular}

\begin{tablenotes}[flushleft]
\footnotesize
\item Designs of order three. Each row uses the same 200 samples for all four fits.
\end{tablenotes}
\end{threeparttable}
\end{table}

\subsection{Confidence sets under weak identification}

At order three with the lower information level and known transitions, the 95\% AR confidence set covers the true discount factor in 93.5\% of samples at 10,000 paths. Its median length is 0.96, the entire search interval, and 94.5\% of sets touch at least one endpoint. The AR point estimate is on the boundary in 64\% of these draws. Even at one million paths, the median set length is 0.74.

\begin{figure}[!htbp]
\centering
\includegraphics[width=.97\linewidth]{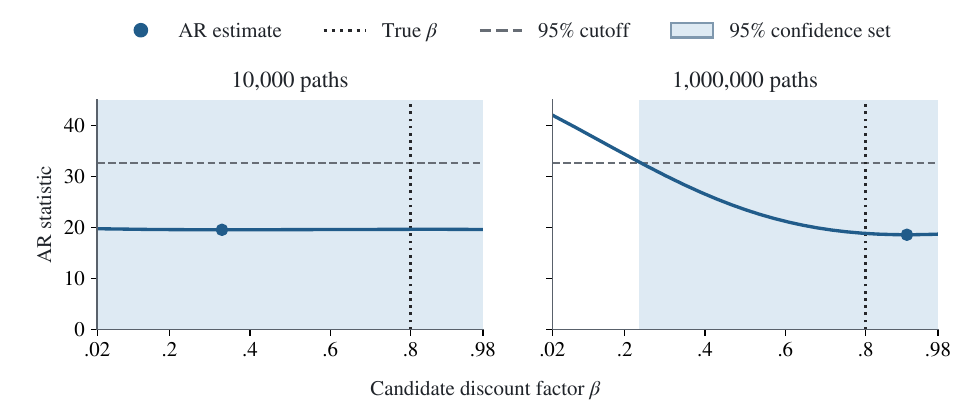}
\caption{AR confidence sets at order three under weak identification}
\label{fig:weak-identification}
\exhibitnote{
Each panel shows one of 200 draws at the indicated sample size, selected for proximity to the median length of the confidence set. The true discount factor is 0.80.
}
\end{figure}

Figure~\ref{fig:weak-identification} shows a typical draw at each sample size. The dot marks the estimate and the shaded region shows the candidates that pass the test. The example with 10,000 paths accepts the entire interval. The example with one million paths rules out values below about 0.24.

\FloatBarrier

\subsection{Comparison with Tsubota's likelihood}
\label{sec:compare-present-bias}

We use the design of order three at the higher information level and set the factor for present bias to $\alpha=1$ or $\alpha=0.70$. The discount factor for the long run is $\delta=0.80$ in both cases. We keep the flow payoffs, transitions and terminal values fixed and simulate 100 samples at each of the same 21 sizes. On each draw we fit AR, geometric MLE and our adaptation of Tsubota's MLE, which estimates $\alpha$. When the data exhibit present bias, we compare the discount factor from each geometric fit with the true $\delta$.

\begin{figure}[!htbp]
\centering
\includegraphics[width=.97\linewidth]{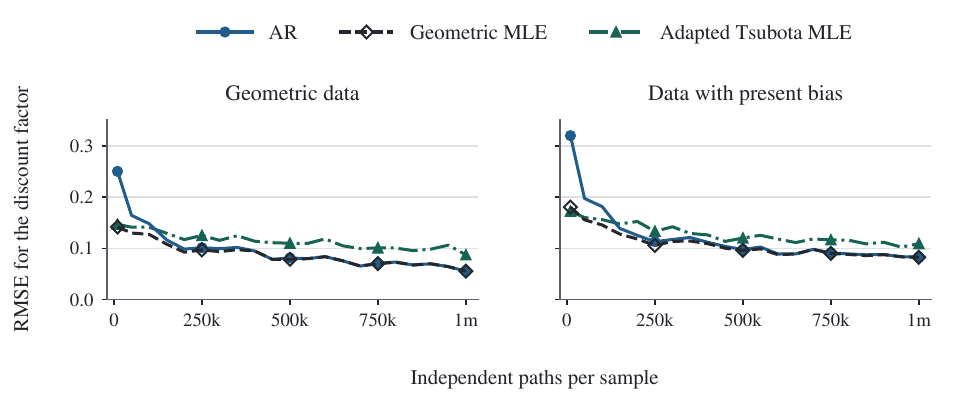}
\caption{RMSE for the discount factor with and without present bias}
\label{fig:present-bias}
\exhibitnote{
Each point uses 100 matched samples. The geometric data have $\alpha=1$ and the data with present bias have $\alpha=0.70$. The adapted Tsubota fit estimates $\alpha$ in both panels.
}
\end{figure}

When the data are geometric, AR and geometric MLE both reach an RMSE of 0.056 at one million paths, compared with 0.088 for the adapted Tsubota fit. When the data exhibit present bias, the adapted fit has less absolute bias than geometric MLE at that size, 0.034 compared with 0.049. Its greater sampling dispersion raises RMSE to 0.109, compared with 0.083 for both AR and geometric MLE. Figure~\ref{fig:present-bias} shows all three methods across the grid.

\FloatBarrier

\section{Conclusion}

A change in the probability of repair at a fixed machine condition reveals a change in the value of repair relative to keeping the machine. When flow payoffs and transitions are time invariant, the law of motion for those values determines how many later dates are needed to eliminate them. With the boundary restrictions, the resulting recurrence characterizes which discount factors can reproduce the observed choices.

With known transitions and regular identification, weighting the full set of restrictions by the inverse of their covariance matrix gives the same asymptotic variance as geometric MLE. In the simulations, a longer recurrence loses precision when it drops equations near the end of the window. Adding them back recovers the shortest fit. Estimating transitions can change how the two estimators compare at the sample sizes we study. The confidence sets remain wide when choices contain little information about discounting. When the data exhibit present bias, estimating the additional parameter reduces the absolute bias of the discount factor at one million paths and raises its RMSE.

\FloatBarrier
{\small\setstretch{1.05}\setlength{\bibsep}{4pt}\interlinepenalty=10000
}

\clearpage
\appendix
\setstretch{1.06}
\numberwithin{equation}{section}
\numberwithin{table}{section}
\numberwithin{figure}{section}
\section{Additional formulas and regularity conditions}\label{app:model}
\subsection{Relative values}
Let $m=K-1$. Write $R=[-\bm 1_m\ I_m]$ for the matrix that subtracts the first component and $E=[0_{1\times m};I_m]$ for the matrix that inserts a zero as the first component. Thus $v_t=RV_t$, $D=(P_1-P_0)E$, $P=RP_0E$, and $h_t=-R\log(1-p_t)$. We write $\Omat_j$ for the matrix stacking the $j$ blocks $D,DP,\ldots,DP^{j-1}$ vertically.

\subsection{Recurrence coefficients and special cases}
Given the matrices in \eqref{eq:C}, the coefficient of each power of the discount factor is
\begin{align}
B_j&=DP^{j-1}-\sum_{k=1}^{j-1}C_{q-k}DP^{j-k-1},\\
a_{j,t}&=C_{q-j}z_{t+j}+B_jg_{t+j}.\label{eq:a}
\end{align}
When $q=1$ and $D$ has full column rank, $C_0=DPD^+$ and $B_1=D$. The recurrence is $z_t=\beta_0(DPD^+z_{t+1}+Dg_{t+1})$. When $D$ lacks full column rank, a matrix $C_0$ satisfying $DP=C_0D$ gives the recurrence of order one in Section~\ref{sec:two-step}.

For $m=2$, the characteristic identity $P^2-\tau P+\delta I=0$, with $\tau=\operatorname{tr}(P)$ and $\delta=\det(P)$, supplies $C_1=\tau I_K$ and $C_0=-\delta I_K$. Consequently,
\begin{align}
z_t={}&\beta_0\big(\tau z_{t+1}+Dg_{t+1}\big)\nonumber\\
&+\beta_0^2\big(-\delta z_{t+2}+(DP-\tau D)g_{t+2}\big).\label{eq:ar2}
\end{align}
The characteristic identity eliminates the unknown values without inverting $D$. Both terms use powers of the same $\beta_0$.
\subsection{Tests at the boundary and across the window}\label{app:boundary}
Equation~\eqref{eq:y} requires $z_t\in\col(D)$ at every date. The joint test also checks the way those outputs are linked. For $i=0,\ldots,n-1$, define
\begin{equation}
F_i(\beta)=\beta^{n-1-i}\left[z_{n-i}-\sum_{l=1}^i\beta^lDP^{l-1}g_{n-i+l}\right],\label{eq:F}
\end{equation}
with an empty sum equal to zero, and stack $F=(F_0^\top,\ldots,F_{n-1}^\top)^\top$. The same formula for $i=0,\ldots,q-1$ defines $F^{(q)}$ (the common factor $\beta^{n-1}$ can equivalently be replaced by $\beta^{q-1}$). Let $\Pi_j=\Omat_j\Omat_j^+$.

The superscript $+$ denotes the Moore--Penrose inverse. The boundary test in Theorem~\ref{thm:ident} is $(I-\Pi_q)F^{(q)}(\beta)=0$. The test across the window applies even when the data are too short for a recurrence.
\begin{equation}
\boxed{\mathcal I=\{\beta\in\B:(I-\Pi_n)F(\beta)=0\}.}\label{eq:sharp}
\end{equation}
\subsection{Statistical assumptions and consistency}\label{app:statistics}
Let $\eta$ collect the population logits and any free transition probabilities. Write $m_q(\beta,\eta)$ for the population moment vector and $\hat m_q(\beta)=m_q(\beta,\hat\eta)$ for its sample counterpart. The number of independent paths or recurrent cycles is $M$. Hold $H$ and $q$ fixed. For estimation, take $\mathcal B$ to be compact. For identification, it may be the entire open unit interval.

\begin{enumerate}[label=(\roman*),leftmargin=*,nosep]
\item $\hat\eta\to_p\eta_0$. Choice probabilities stay away from zero and one, and the states and actions used to estimate transition rows have positive support.
\item $m_q$ is continuous in $(\beta,\eta)$ locally uniformly over $\B$. The order is fixed and the rank parameterization is stable.
\item $\hat W_q$ converges uniformly to a continuous symmetric nonnegative weight $W_q(\beta)$.
\item The population criterion has a unique minimizer $\beta_0$. A singleton root set and positive definite weights suffice.
\end{enumerate}

With known transitions, continuity of the local moments holds directly. With estimated transitions, a sample pseudoinverse can be discontinuous when rank changes. Available representations include a parameterization with fixed rank, a recurrence based on the characteristic polynomial, and a test across the window when its terminal matrix has full column rank. The experiment with estimated transitions uses the last option in the design of order three.

For asymptotic normality, also assume
\begin{enumerate}[label=(\roman*),start=5,leftmargin=*,nosep]
\item $\sqrt M(\hat\eta-\eta_0)\Rightarrow N(0,\Sigma_\eta)$.
\item $m_q$ is twice continuously differentiable near $(\beta_0,\eta_0)$. The weight and its derivative converge locally uniformly and are bounded. A fixed weight needs no derivative condition.
\item $G=\partial_\beta m_q(\beta_0,\eta_0)$ satisfies $G^\top W G>0$ with $W=W_q(\beta_0)$, and $\beta_0$ is interior.
\end{enumerate}
Redundant moments can make $\Omega=M_q(\beta_0)\Sigma_\eta M_q(\beta_0)^\top$ singular. In that case, use a stable basis or a pseudoinverse with constant rank and retain condition (iv).

Here $M_q(\beta)=\partial_\eta m_q(\beta,\eta_0)$ and $\Omega_q(\beta)=M_q(\beta)\Sigma_\eta M_q(\beta)^\top$. The sample covariance substitutes the corresponding estimates into this expression.
\subsection{Dependent sampling and inference by moment tests}
For independent panels, each observation is a full state and action path. Choices within a path may be dependent. The cell and transition probabilities are ratios of average indicator counts, so a central limit theorem across panels and the delta method give condition (v). For a single recurrent sequence, group observations into cycles with the same $H$ policy functions, one for each phase, and a fixed transition kernel. Suppose the cycle process is stationary and mixes enough for a multivariate central limit theorem. Let $X_j$ contain the counts in cycle $j$ and let $f$ map their mean to $\eta$.
\begin{equation}
\Sigma_\eta=f'(\E X_j)\left[\sum_{\ell=-\infty}^{\infty}\operatorname{Cov}(X_j,X_{j-\ell})\right] f'(\E X_j)^\top.\label{eq:hac}
\end{equation}
The variance formula remains valid under standard conditions for consistent estimation of the long run covariance. A periodic Markov process on finitely many states with positive transition probabilities has a geometrically ergodic cycle chain, so the initial condition has a negligible asymptotic effect. Repeated phases provide multiple observations of the choice probabilities for each phase. The independence assumption on private shocks still applies.

In our logit experiments with known transitions, the action residual $A_t-p_t(S_t)$ has conditional mean zero given phase and state. The cell logit influence scores are therefore martingale differences. Their covariance across dates and states is zero despite dependence in the state process. For $n_{ts}$ visits to a cell, the estimated logit covariance is diagonal with entries $[n_{ts}\hat p_{ts}(1-\hat p_{ts})]^{-1}$ on the unnormalized estimator scale. Differencing still uses the same logits in multiple equations, so the moment covariance is dense.

When identification is weak, Wald intervals may be misleading. A complementary confidence set inverts the pointwise statistic
\begin{equation}
\mathcal C_{1-\alpha}=\{\beta\in\B:M\hat m_q(\beta)^\top\hat\Omega_q(\beta)^+\hat m_q(\beta)\le\chi^2_{r,1-\alpha}\},\label{eq:set}
\end{equation}
where $r$ is the stable covariance rank at the candidate. At the true parameter, (v), differentiability and consistent covariance estimation imply a $\chi^2_r$ limit without requiring $G\ne0$. The null limit is pointwise at the candidate covariance rank. The set may be large, disconnected or empty.

\section{Proof of Theorem \ref{thm:ident}}\label{app:proof1}
\begin{proof}
\emph{Step 1. Reduction.} Since $RE=I_m$ and $ER=I_K-\bm 1_K e_1^\top$, write $V=Ev+V(1)\bm 1_K$. The identities $(P_1-P_0)\bm 1_K=0$ and $RP_0\bm 1_K=0$ give \eqref{eq:reduced} directly from \eqref{eq:logit}--\eqref{eq:bell}. Time differencing yields \eqref{eq:ss}. The definition $y_t=\beta w_{t+1}$ then gives \eqref{eq:y} for $1\le t<n$. No boundary value is used in the derivation.

\emph{Step 2. Elimination.} Fix a candidate $\beta>0$. Iterating \eqref{eq:y} $q$ times gives
\begin{equation}
z_t=\sum_{l=1}^{q}\beta^lDP^{l-1}g_{t+l}+\beta^q DP^q y_{t+q}.\label{eq:iterate}
\end{equation}
For each $j=0,\ldots,q-1$, the same identity, now beginning at $t+q-j$ and iterated $j$ times, implies
\begin{equation}
DP^j y_{t+q}=\beta^{-j}\left[z_{t+q-j}-\sum_{r=1}^j\beta^rDP^{r-1}g_{t+q-j+r}\right].\label{eq:sub}
\end{equation}
By closure, $DP^q=\sum_{j=0}^{q-1}C_jDP^j$. Substituting \eqref{eq:sub} into \eqref{eq:iterate}, and collecting the terms associated with $z_{t+l}$ and $g_{t+l}$ at displacement $l$, yields
\begin{align*}
z_t={}&\sum_{l=1}^q \beta^l C_{q-l}z_{t+l}\\
&+\sum_{l=1}^q\beta^l\left(DP^{l-1}-\sum_{k=1}^{l-1}C_{q-k}DP^{l-k-1}\right)g_{t+l}.
\end{align*}
This proves \eqref{eq:ar} with \eqref{eq:a}. Let $A$ have columns given by stacking $a_{j,t}$ over states and dates. If $A$ has full column rank, two candidates satisfying the recurrence must have identical vectors of powers, hence identical first components. If there is only one common root, every rationalizing candidate must equal it. Existence is guaranteed by the maintained true model.

\emph{Step 3. Necessity of compatibility across the window.} Iterate \eqref{eq:y} from $n-i$ to $n$. Then
\[
z_{n-i}-\sum_{l=1}^i\beta^l DP^{l-1}g_{n-i+l}=\beta^i DP^i y_n.
\]
Multiplication by $\beta^{n-1-i}$ yields $F_i(\beta)=\beta^{n-1} DP^i y_n$. Stacking all $i$ proves $F(\beta)\in\col(\Omat_n)$ and hence the necessity of \eqref{eq:sharp}.

\emph{Step 4. Sufficiency of compatibility across the window.} Suppose \eqref{eq:sharp} holds. Since $\beta>0$, choose $y_n$ with $F(\beta)=\beta^{n-1}\Omat_n y_n$. Define $y_t$ backward for $t=n-1,\ldots,1$ by $y_t=\beta g_{t+1}+\beta P y_{t+1}$. The iteration identity and the defining equations for $F_i$ guarantee $z_t=Dy_t$ at every $t$. We next construct primitive utility and value vectors that rationalize the original CCPs. Let $w_{t+1}=y_t/\beta$ for $t=1,\ldots,n$, and $w_1=g_1+\beta P w_2$. Choose arbitrary $v_{H+1}\in\R^m$ and set $v_t=v_{t+1}+w_t$ backward for $t=H,\ldots,1$. Use the last date to define the two payoff differences.
\begin{align*}
b&=L_H-\beta Dv_{H+1},\\
c&=v_H-h_H-\beta Pv_{H+1}.
\end{align*}
The first expression $L_t-\beta Dv_{t+1}$ is time invariant because its difference is $z_t-\beta Dw_{t+1}=0$. The second expression $v_t-h_t-\beta Pv_{t+1}$ is time invariant because its difference is $w_t-g_t-\beta Pw_{t+1}=0$. Thus \eqref{eq:reduced} holds at every date. Set $u_0=Ec$ and $u_1=u_0+b$. Choose the terminal level of $V_{H+1}$ arbitrarily with relative component $v_{H+1}$. Recover full $V_t$ backward by \eqref{eq:bell} with observed $p_t$. Applying $R$ to that recursion shows by induction that its relative components equal the constructed $v_t$. Consequently $Q_{1,t}-Q_{0,t}=L_t$. Also $V_t-Q_{0,t}=-\log(1-p_t)=\log(1+e^{L_t})$, so $V_t=\log(e^{Q_{0,t}}+e^{Q_{1,t}})$. The resulting constant flow payoffs and the terminal value therefore rationalize all observed CCPs. This proves sharpness.

\emph{Step 5. Equivalence to the recurrence and boundary test.} Necessity follows from Steps 2--3. For sufficiency, suppose the local equations hold and $F^{(q)}(\beta)\in\col(\Omat_q)$. The boundary condition gives a vector $y_n$ that matches the last $q$ outputs under \eqref{eq:y}. Define all earlier $y_t$ by backward recursion and denote their implied outputs by $\tilde z_t$. The last $q$ values of $\tilde z$ equal those of $z$. By Step 2, $\tilde z$ satisfies the same recurrence \eqref{eq:ar} with the same fixed $g$. Induction backward on $t$ therefore gives $\tilde z_t=z_t$ for every preceding $t$. Step 4 supplies a structural rationalization. Without the boundary condition, the recurrence can admit trajectories that no terminal vector can generate.

\emph{Step 6. Existence of finite closure.} The matrix $P$ satisfies its characteristic polynomial of degree $m$, so $DP^m$ belongs to the row span of $\Omat_m$. If equality of successive ranks holds at some order $q$, $\row(\Omat_q)$ is invariant under right multiplication by $P$ because multiplying each generating block by $P$ stays in the same span. Closure therefore persists. If $D=0$, any compatible $z$ must be zero and the sharp restriction holds for all $\beta$. Full column rank of $D$ ensures closure at order one. Closure at order one also holds if $D$ lacks full column rank but its row space is invariant under right multiplication by $P$.
\end{proof}
\section{Proof of Theorem \ref{thm:cons}}
\begin{proof}
Write $Q_M(\beta)=m_q(\beta,\hat\eta)^\top\hat W_q(\beta)m_q(\beta,\hat\eta)$ and its limit $Q(\beta)$ with $\eta_0$ and $W_q(\beta)$. Compactness of $\B$ and local joint continuity imply $\sup_{\beta\in\B}\|m_q(\beta,\hat\eta)-m_q(\beta,\eta_0)\|=o_p(1)$. The population moments are bounded. Uniform convergence of weights and a bound on the difference of the quadratic forms then imply $\sup_{\beta\in\B}|Q_M(\beta)-Q(\beta)|=o_p(1)$. Since $Q$ is continuous with a unique minimizer $\beta_0$, for every $\varepsilon>0$ its minimum over $\{\beta:|\beta-\beta_0|\ge\varepsilon\}$ exceeds $Q(\beta_0)$ by a strictly positive amount (if the set is nonempty). Uniform convergence rules out a sample minimizer in that set with probability tending to one. The same reasoning with the distance to the compact minimizing set instead of the distance to $\beta_0$ gives convergence to the minimizing set for an approximate minimizer with objective error $o_p(1)$.
\end{proof}
\section{Proof of Theorem \ref{thm:normal}}
\begin{proof}
For brevity denote $m=m_q$, $G=\partial_\beta m(\beta_0,\eta_0)$ and $J=\partial_\eta m(\beta_0,\eta_0)$. Let $\Omega$ be the limiting covariance of the scaled moments and $W$ the limiting weight. The central limit theorem for the first stage and differentiability give
\begin{equation}
\sqrt M\,m(\beta_0,\hat\eta)=J\sqrt M(\hat\eta-\eta_0)+o_p(1)\Rightarrow N(0,\Omega).
\end{equation}
Since the true parameter is interior and the estimator is consistent, an exact minimizer satisfies the condition that the derivative vanishes with probability tending to one. For fixed or continuously updated weights it is
\[
0=2\partial_\beta\hat m(\hat\beta)^\top\hat W(\hat\beta)\hat m(\hat\beta)
 +\hat m(\hat\beta)^\top\partial_\beta\hat W(\hat\beta)\hat m(\hat\beta),
\]
where the last term is zero for fixed weights. Write $\Delta=\hat\beta-\beta_0=o_p(1)$. Smoothness and the convergence rate in the first stage yield
\[
\hat m(\hat\beta)=m(\beta_0,\hat\eta)+G\Delta+o_p(|\Delta|),
\quad \partial_\beta\hat m(\hat\beta)=G+o_p(1).
\]
The weight derivative term is $O_p((M^{-1/2}+|\Delta|)^2)=o_p(M^{-1/2}+|\Delta|)$. Uniform convergence of the weight gives
\[
0=G^\top W m(\beta_0,\hat\eta)+G^\top WG\,\Delta+o_p(M^{-1/2}+|\Delta|).
\]
Because $G^\top WG>0$, this first implies $\Delta=O_p(M^{-1/2})$ and then
\[
\sqrt M\,\Delta=-(G^\top WG)^{-1}G^\top WJ\sqrt M(\hat\eta-\eta_0)+o_p(1).
\]
The limit for the first stage gives
\begin{equation}
\sqrt M(\hat\beta-\beta_0)\Rightarrow
N\!\left(0,\frac{G^\top W\Omega W G}{(G^\top W G)^2}\right).
\label{eq:variance}
\end{equation}
The variance is $(G^\top\Omega^{-1}G)^{-1}$ when the weight is the inverse of the covariance matrix. The same formula uses the Moore--Penrose inverse on a stable covariance range when $G^\top\Omega^+G>0$.
\end{proof}
\section{Proof of Proposition \ref{prop:efficiency}}\label{app:efficiency}
\begin{proof}
Write $L$ for the vector of the logits for all states and dates. At a fixed $\beta>0$, the map from $L$ to $(L_H,F)$ in \eqref{eq:F} is smooth and invertible. To see this, start with $L_H$ and recover the preceding logits one date at a time. In each equation for $F_i$, the current logit enters $z_{n-i}$ with coefficient $\beta^{n-1-i}$, and the inputs in the sum involve only later logits. The coefficient is nonzero.

By \eqref{eq:sharp}, rationalizable logits are exactly those with $F$ in the column space of $\Omat_n$. At fixed $\beta$, we can therefore parameterize the nuisance part of the likelihood by $L_H$ and the coordinates of $F$ in that space. Project $F$ onto an orthonormal basis of its orthogonal complement to obtain the independent moments $m(\beta,L)$. If $J$ is their derivative with respect to $L$, then $J$ has full row rank. The nuisance directions in logit space are exactly $\ker J$.

Let $\Sigma$ be the asymptotic covariance per path of the unrestricted logit estimates. Its inverse is the conditional likelihood information for $L$. Indeed, the action scores at successive dates are martingale differences, and the diagonal information entry in each cell is its visit probability times its Bernoulli variance. Positive visit probabilities and interior choice probabilities give $\Sigma>0$.

Let $d$ denote the derivative of the structural logits with respect to $\beta$ at fixed nuisance coordinates. Differentiating $m(\beta,L)=0$ along this path gives $Jd=-G$, where $G=\partial_\beta m$ holds $L$ fixed. Profiling out the nuisance directions gives information
\[
\min_{v:\,Jv=0}(d-v)^\top\Sigma^{-1}(d-v)
= (Jd)^\top(J\Sigma J^\top)^{-1}Jd
=G^\top(J\Sigma J^\top)^{-1}G.
\]
The first equality follows by minimizing over $d-v$ subject to $J(d-v)=Jd$. The final expression is the inverse of the optimally weighted variance in \eqref{eq:variance}. Smooth invertible changes of moment coordinates at the population restriction leave this expression unchanged.
\end{proof}

\section{Derivation of the conditional mean benchmark}\label{app:cm}
For the supplementary estimator based on direct conditional means, normalize the payoff of the reference action to $u_0=0$ and write $b=u_1$. The expected normalized shock of the chosen action gives the binary entropy correction
\[
e_j(s)=-p_j(s)\log p_j(s)-(1-p_j(s))\log(1-p_j(s)).
\]
Under the optimal logit policy, the integrated value at date $j$ is the expected discounted stream of $p_j(s)b(s)+e_j(s)$ and the terminal value. Let $Z_{tj}$ be the $K\times K$ matrix whose row $s$ is the difference between the conditional distributions of $S_j$ given $S_t=s,A_t=1$ and $S_t=s,A_t=0$, respectively. The log odds thus satisfy
\begin{equation}
L_t=b+\sum_{j=t+1}^H\beta^{j-t}Z_{tj}\big[\operatorname{diag}(p_j)b+e_j\big]+\beta^{H+1-t}Z_{t,H+1}E\gamma,
\end{equation}
where $E\gamma$ is the terminal vector after removing its constant. This expression iterates the integrated Bellman equation under the optimal policy. The estimator uses cell frequencies for $p_j$ and conditional frequencies of future states for $Z_{tj}$. For each candidate $\beta$, the equation is linear in $(b,\gamma)$ and is fit by weighted least squares using empirical logit precision. Then we minimize the profiled criterion over $\beta$ using the same grid and scalar refinement approach as CUE. The calculation uses estimated conditional means in place of transition probabilities. The nuisance coefficients are unrestricted, and the weights do not fully account for estimation error in $Z_{tj}$. This estimator is an adaptation of the conditional mean approach.

\section{Simulation design and additional results}
\label{app:simulations}

\subsection{Worked example and nested observation windows}
We generate Table~\ref{tab:two-state-choices} by backward induction with $u_0=(0,0)^\top$, $u_1=(-1,0.4)^\top$, $V_4=(2,0)^\top$, and $\beta_0=0.8$. From good and worn condition, respectively, the probabilities of a worn machine next period are $(0.25,0.75)$ under action 0 and $(0.05,0.55)$ under action 1. The calculation in \eqref{eq:two-state-beta} uses the resulting probabilities and transitions to recover $0.8$ in both states. The flow payoffs and $V_4$ enter only when generating the probabilities.

To compare observation windows, retain the first $H'$ dates of each population choice path, with $H'=2,\ldots,7$. In each design, $D$ has rank one and the components orthogonal to its output direction are zero. Project onto that direction. The first $q$ rows of the observability system are independent, and multiplying by nonzero discount powers leaves their rank intact. When $H'-1\le q$, a free boundary vector can match all projected outputs for any positive discount factor. When $H'-1>q$, it can match the final $q$ projected outputs, so only the local equations restrict $\beta$. We find the roots in $(0,1)$ of a nonconstant moment polynomial and check each root against the remaining equations. Only $0.8$ passes in every longer window.

\subsection{Matching the amount of information about discounting}
All designs have $H=7$, $K=4$, and $\beta_0=0.8$. Let $e_j$ be coordinate vector $j$. The first row of $P_0$ is $(0.85,0.05,0.05,0.05)$. Row $i+1$ adds $\lambda_i(e_{i+1}-e_1)^\top$ to that first row. Set $D=d v^\top$ with $d=(0.16,-0.035,0.075,-0.035)^\top$, $v=(1,0.8,1.2)^\top$, and $P_1=P_0+[-D\mathbf 1_3\; D]$. The eigenvalue triples are $(0.55,0.55,0.55)$, $(0.20,0.20,0.75)$, and $(0.10,0.47,0.78)$ for minimum orders one, two, and three respectively. The flow payoff for the reference action is zero, and the other payoff is $(-0.7,0.15,-0.25,0.6)^\top$. The direction of the relative terminal payoff is the orthogonal projection of $(2.8,-1.7,2.1)^\top$ onto the row space of $\Omat_3$. Only its multiplier changes within each model to set the information level.

We set the payoff multiplier using the population Fisher information of the conditional choice likelihood, after profiling out the flow payoffs and observable coordinates of the terminal payoff. We start with uniform initial states and obtain the state probabilities at each date from the optimal choice probabilities and the two transition matrices. Let $J$ be the derivative of the logits for each state and date with respect to the discount factor and the nuisance parameters. Weight each row of $J$ by the square root of the state probability times $p(1-p)$. Project the column for the discount factor onto the nuisance columns. The squared norm of the residual gives the information per path. We choose the multiplier to match $3\times10^{-5}$ or $3\times10^{-4}$. The two targets concern the curvature of the profiled likelihood at the true value. The likelihood elsewhere, the state distribution, and the nuisance parameters can differ across designs.

\begin{table}[htbp]
\centering
\caption{Calibration of terminal payoffs}
\label{tab:calibration}
\small\begin{tabular}{rrr}
\toprule
Minimum order & Lower information & Higher information \\
\midrule
1 & 0.548 & 0.764 \\
2 & 0.994 & 1.641 \\
3 & 1.480 & 4.782 \\
\bottomrule
\end{tabular}

\exhibitnote{
Information targets per path are $3\times10^{-5}$ and $3\times10^{-4}$. Entries are the terminal payoff multipliers.
}
\end{table}

We draw counts by state and action using binomial sampling within states and multinomial transitions for each combination of state and action. This yields the same distribution of state and action counts by date as drawing independent paths one at a time. Both actions occur in each cell in the simulated samples. For the population root check, we take the real roots in $(0,1)$ of an informative moment and test each root against every included moment. The only common root is $0.8$ in all six specifications using the shortest recurrence and in the longer recurrences we use.

\subsection{Why restoring the final equations recovers the shortest fit}
In the simulated designs with rank one, the projected recurrence supplies one scalar equation per date where it applies. In a model with shortest order $r$, the last $r$ scalar outputs can be matched by the terminal state vector after restricting the outputs to the column space of $D$. The $r$ corresponding rows of the observability system are independent. The shortest rule then supplies $n-r$ earlier equations.

A longer order $q$ gives $n-q$ earlier equations. We add the shortest rule at the next $q-r$ dates, where its required later changes are available. For fixed discounting and inputs, either system reconstructs the earlier outputs from the last $r$ outputs by backward substitution. They reconstruct the same sequence. A solution of the shortest rule that satisfies those restrictions is compatible with the structural model and therefore obeys the longer rule. For arbitrary inputs, the projected residuals from either fit are invertible triangular transformations of the deviations of the first $n-r$ outputs from that solution. Both systems also include the same residuals for the restrictions on the outputs.

With known transitions, this change of residual coordinates depends on the candidate discount factor and the transitions. It is fixed with respect to the estimated choice probabilities. If the moment covariance is invertible, applying the same invertible transformation to the moments and their covariance leaves the continuously updated objective unchanged. The restored fit uses the shortest order supplied by the design to construct the extra boundary equations.

\subsection{Estimation and Monte Carlo uncertainty}
We estimate the logits from action counts and use the full moment covariance induced by those logits. The scalar search compares the values at 61 grid points on $[0.02,0.98]$. It refines each detected local minimum and the intervals next to both endpoints.

For likelihood, we fit weighted deviance residuals using the Bellman recursion and analytic derivatives. We initialize discounting at $0.10$, $0.35$, $0.75$, and $0.95$ and retain the fit with the lowest objective. The flow and terminal payoff coordinates lie in $[-30,30]$. The benchmark estimates a free payoff in each state with the payoff of the reference action normalized to zero and fits the observable terminal coordinates. We can set the reference payoff to zero by subtracting $(I-\beta P_0)^{-1}u_0$ from the value vectors and adjusting the other action's payoff and the terminal value. All selected likelihood fits in the experiment with matched information levels met the optimizer's stopping rules.

We obtain the Monte Carlo standard error of the RMSE gap by applying the delta method to the difference between the square roots of the two mean squared errors. We pair the squared errors by replication. The replication files also record bias, boundary events, and whether the pointwise 95\% moment test accepts the true discount factor.

\subsection{Estimated transitions and confidence sets}

For the design of order three, we replay the same 200 draws at each information level and at 10,000, 100,000 and one million paths. The estimated transition rows are pooled frequencies of next states conditional on the current state and action. The centered action indicators and the centered indicators for the next state are martingale differences in the order in which they are observed. The covariance between the estimated logits and the transition frequencies is therefore zero to first order. We use the binomial logit covariance and the multinomial covariance for each row, with the first entry of each transition row recovered from the other three.

For the AR fits with estimated transitions, we stack the equations across the window in Appendix~\ref{app:boundary} and use a QR decomposition to project out the three terminal coordinates. The stacked terminal matrix has full column rank at the population transitions in this design, even though $D$ alone has rank one. We keep the three coordinates throughout. For the covariance, we differentiate the residuals across the window with respect to the logits and transition probabilities, evaluate the derivative at the terminal vector fitted by least squares at each candidate, and apply the same projection. The likelihood fit with estimated transitions uses the same bounds and adds one start at the AR estimate to the four fixed starts. Its local standard errors propagate the transition covariance through the derivative of the fitted choice equations.

At each of the three sample sizes at the lower information level, we invert the test with known transitions in \eqref{eq:set}. We check 481 equally spaced candidates on $[0.02,0.98]$ and refine threshold crossings between adjacent candidates. Set length is the sum of the lengths of all accepted intervals, with zero for an empty set. An endpoint counts as touched if it belongs to the set. For each panel in Figure~\ref{fig:weak-identification}, we choose the draw closest to the median set length. Ties go to the draw whose statistic at $\beta_0$ is closest to the median across draws at that size, then to the earliest draw.

\subsection{Likelihood comparison with present bias}
\label{app:present-bias}

We use the backward recursion of \citet[][Section 4]{tsubota2026} with the payoff for action 0 set to zero and an unrestricted terminal vector. Let $\Delta P=P_1-P_0$. At a trial vector with discount factor $\delta$ for the long run and factor $\alpha$ for present bias, the log odds and continuation value satisfy
\begin{align}
 L_t &= u_1+\alpha\delta\Delta P V_{t+1},\\
 V_t &= \delta P_0V_{t+1}-\log(1-p_t)
       +(1-\alpha)\delta\,\operatorname{diag}(p_t)\Delta P V_{t+1}.
\end{align}
The last term adjusts the value of future choices to reflect how they are discounted in the long run. Setting $\alpha=1$ removes it and gives the geometric recursion in Section~\ref{sec:model}. The trial vectors include a payoff for action 1 in every state and three relative terminal coordinates. All fits use the same true transitions. The MLE for the model with present bias searches over $\alpha\in[0.02,1]$ and $\delta\in[0.02,0.98]$. We use four starts, including the geometric fit at $\alpha=1$, and retain the lowest deviance. The population logit Jacobian has full column rank at both generating parameter vectors. The exercise compares likelihood fits on our calibrated design with an unobserved terminal value, a different setup from Tsubota's simulations with a zero terminal value.

\subsection{Likelihood initialization and timing protocol}
\label{app:likelihood-starts}
For timing, we rerun the first 100 draws per calibration at 10,000, 100,000, and one million paths. The 18 cells run on the same machine, serially and with one BLAS thread. Within a replication we rotate the order of the four timing strategies. The search with four starts initializes discounting at $0.10$, $0.35$, $0.75$, and $0.95$. At each fixed start, the initial flow utility vector is the vector of logits from the last date and the terminal coordinates start at zero.

For the AR start, fix the discount factor at the continuously updated estimate from minimum distance. With the reference payoff normalized to zero, iterate $v_t=h_t+\beta Pv_{t+1}$ backward. Given the estimated $h_t$, substituting this recursion into $L_t=b+\beta Dv_{t+1}$ gives a linear equation in the differences in flow payoffs $b$ and the observable coordinates of $v_{H+1}$. We fit that equation by least squares, weighting the logit for each state and date by its estimated precision, and project the starting coordinates into the likelihood search box. The joint optimization then updates $\beta$, $b$ and the terminal coordinates. The timed AR fit includes constructing the moment covariance and optimizing the minimum distance criterion. We record the stopping flag and evaluate the exact conditional choice log likelihood at the solution from every fit. The fit with four starts is the numerical benchmark.

Following the computational comparison of \citet{buntingura2025}, we record both computational cost and the value of a common target objective. We compare the best likelihood fit from four fixed starts with one fixed start at $\beta=0.75$, one AR start, and the better fit from the AR and fixed starts. All strategies use the same likelihood, bounds, analytic derivatives and stopping tolerances.

\begin{table}[tbp]
\centering
\caption{Likelihood fit and computation by choice of starting values}
\label{tab:likelihood-starts}
\begin{tabular}{lrr}
\toprule
Starting values & Speedup & Lower likelihood \\
\midrule
Four fixed starts & 1.00 & 0 \\
One fixed start & 4.83 & 13 \\
One AR start & 5.61 & 14 \\
AR and fixed starts & 2.70 & 0 \\
\bottomrule
\end{tabular}

\exhibitnote{
Results pool 1,800 matched samples. Speedup is the median of the ratios of benchmark runtime to the runtime of each strategy on the same draw. All runtimes include the cost of constructing starting values. A lower likelihood is more than 0.01 below the benchmark with four fixed starts.
}
\end{table}
The fixed start gives a lower likelihood on 13 draws, and the AR start does so on 14. The 14 AR misses all occur in the design of order three at the lower information level at the two smaller sample sizes. The two sets of misses are disjoint. Taking the better of the two fits attains the benchmark likelihood within tolerance on all draws. Part of the time gain comes from running fewer optimizations. The AR start supplies the benchmark fit on the draws where the fixed start alone misses. All selected fits met the numerical stopping criteria.

\subsection{Original payoff design}
\begin{figure}[htbp]
\centering
\includegraphics[width=.94\linewidth]{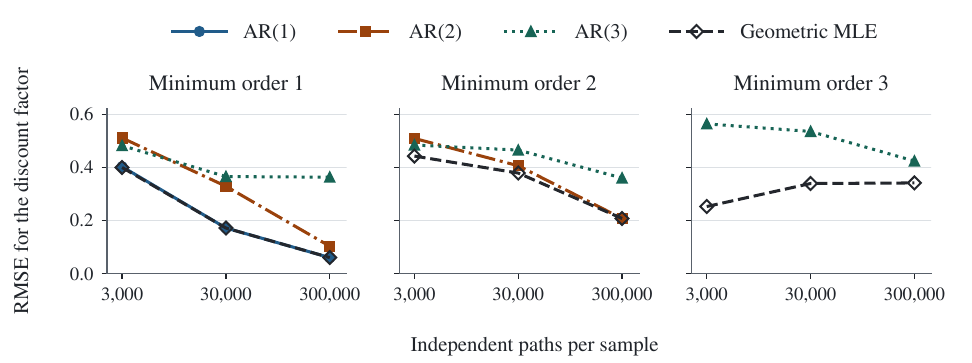}
\caption{Error in the discount factor with the original terminal payoffs}
\label{fig:original}
\exhibitnote{
Each point uses 100 matched samples. A curve appears only at an admissible order. The horizontal axis is logarithmic.
}
\end{figure}

In the original specification, every terminal payoff multiplier is one, so information differs across models. With these payoffs, the information per path at order three is about $1/88$ of the information at order one. At order three and 300,000 paths, AR and likelihood have RMSE 0.4241 and 0.3418, respectively. Figure~\ref{fig:original} reports the results at all sample sizes. The design of order one also shows a large loss when we switch from the shortest recurrence to the local recurrence of order three. The experiment in the main text matches information across models and fits the longer recurrence with the missing conditions restored.

\FloatBarrier

\subsection{One decision maker across repeated contracts}
The agent optimizes within contracts lasting seven dates. The state evolves under the chosen action and carries into the next contract. The choice rules for each phase are the same across contracts. Positive transitions make the chain at the level of contracts ergodic. For each design and each of three lengths, we simulate 50 independent trajectories and fit the shortest autoregression and likelihood to the same counts. In the longest trajectories, AR is close to likelihood at orders one and two. At order three the gap remains substantial (Table~\ref{tab:series}). We use the original terminal payoffs and supply the transitions. Repeated phases identify the choice probabilities.

\begin{table}[htbp]
\centering
\caption{Estimates from a single recurrent trajectory}
\label{tab:series}
\small\begin{tabular}{rrrr}
\toprule
Minimum order & Cycles & AR RMSE & MLE RMSE \\
\midrule
1 & 3,000 & 0.4078 & 0.4047 \\
1 & 30,000 & 0.1728 & 0.1728 \\
1 & 300,000 & 0.0597 & 0.0597 \\
2 & 3,000 & 0.5348 & 0.4460 \\
2 & 30,000 & 0.3396 & 0.3268 \\
2 & 300,000 & 0.2273 & 0.2274 \\
3 & 3,000 & 0.5343 & 0.2461 \\
3 & 30,000 & 0.5236 & 0.2903 \\
3 & 300,000 & 0.4353 & 0.3207 \\
\bottomrule
\end{tabular}

\exhibitnote{
Each cycle contains seven decisions. Each row uses 50 trajectories.
}
\end{table}
\FloatBarrier

\subsection{Direct conditional means with unobserved transitions}
At each of two sample sizes, we draw 50 more samples of individual paths from the original design of order three. The direct estimator uses conditional means of future states, including the terminal state. The other fits use known transitions. The adaptation based on weighted least squares in Appendix~\ref{app:cm} has RMSE 0.584 at 3,000 paths and 0.585 at 30,000. Comparing with \citet{chou2025} would require common information about the unknown transitions.

\end{document}